\documentclass[twocolumn]{article}
\usepackage{authblk}
\usepackage{amsmath,graphicx,booktabs,hyperref}
\usepackage{caption}
\title{MitoDetect++: A Domain-Robust Pipeline for Mitosis Detection and Atypical Subtyping}
\author[1]{Esha Sadia Nasir}
\author[1]{Jiaqi Lv}
\author[1]{Mostafa Jahanifar}
\author[1]{Shan E Ahmed Raza}
\affil[1]{Tissue Image Analytics Centre, University of Warwick}
\date{}
\begin{document}
\twocolumn[
\maketitle
\bibliographystyle{plain}

\begin{abstract}
Automated detection and classification of mitotic figures especially distinguishing atypical from normal remain critical challenges in computational pathology. We present \textbf{MitoDetect++}, a unified deep learning pipeline designed for the MIDOG 2025 challenge, addressing both mitosis detection and atypical mitosis classification. For detection (Track 1), we employ a U-Net-based encoder-decoder architecture with EfficientNetV2-L as the backbone, enhanced with attention modules, and trained via combined segmentation losses. For classification (Track 2), we leverage the Virchow2 vision transformer, fine-tuned efficiently using Low-Rank Adaptation (LoRA) to minimize resource consumption. To improve generalization and mitigate domain shifts, we integrate strong augmentations, focal loss, and group-aware stratified 5-fold cross-validation. At inference, we deploy test-time augmentation (TTA) to boost robustness. Our method achieves a balanced accuracy of \textbf{0.892} across validation domains, highlighting its clinical applicability and scalability across tasks.
\end{abstract}

\vspace{0.5cm}
]
\section{Introduction}
Mitosis detection and classification in histopathological images are essential for accurate tumor grading and prognosis assessment. However, these tasks are inherently challenging due to the rarity of mitotic events, inter-observer variability, and limited atypical mitosis examples. The MIDOG 2025 challenge introduces two critical tasks: \textbf{mitosis detection} (Track 1) and \textbf{atypical mitosis classification} (Track 2), both of which require robust, domain-adaptive solutions.

To this end, we introduce \textbf{MitoDetect++}, a dual-stage pipeline tailored to both tasks:

\begin{itemize}
    \item \textbf{Track 1 (Detection)}: We use an attention-augmented encoder-decoder architecture with an EfficientNetV2-L encoder to detect mitotic figures via binary segmentation, treating mitosis centroids as circular disks.
    \item \textbf{Track 2 (Classification)}: We adopt the ViT-based Virchow2 model as a feature extractor, fine-tuned using Low-Rank Adaptation (LoRA) for resource-efficient training.
\end{itemize}

To combat domain variability and class imbalance across both tracks, we apply strong augmentation strategies, focal loss, and a group-aware stratified cross-validation scheme to avoid slide-level data leakage. Inference is further strengthened by  test-time augmentation (TTA).

Through this multi-faceted approach, MitoDetect++ achieves competitive performance while remaining efficient and generalizable key factors for future clinical deployment.
\section{Track 1 (Mitosis Detection)}
\subsection{Model Architecture}
We created an encoder-decoder model with U-Net style skip connections. The encoder we used was EfficientNetV2-L pre-trained on ImageNet, spatial and channel attention modules were added into the decoder blocks.

\subsection{Dataset and Split Strategy}
We first split MIDOG++ dataset~\cite{aubreville_comprehensive_2023} into train and test set, with test set being 10\% of the entire dataset, split is stratified based on tissue types. We perform 5-fold cross validation on the rest of the training set, stratified also based on tissue types. 

We also sample patches extracted from the CMC dataset~\cite{aubreville_completely_2020}, CCMCT dataset~\cite{CCMCT}, and an mitosis detection dataset published by Mostafa et al~\cite{jahanifar2025mitosis}. during training.

\subsection{Preprocessing}
We extract patches from the datasets. Patch size 512x512 at 0.25 microns per pixel(mpp). We dilate the centroids of mitotic fugues by a diameter of 21 pixels, turning detection into a binary segmentation task.

\subsection{Training Configuration}
\begin{itemize}
    \item AdamW optimiser with initial learning rate $4\times10^{-4}$, weight decay 0.01.
    \item A combination of Jaccard Loss, Dice Loss and Focal Loss.
    \item RandomSampler to keep at least 40\% of patches with mitotic figures per batch of data during training.
    \item Early stopping if validation loss does not decrease in 20 epochs.
\end{itemize}

\subsection{Test Configuration}
The final submitted method used an ensemble of the top 3 model checkpoints from internal 5-fold cross validation. The models’ predictions were averaged.

\section{Track 2 (Mitosis Classification)}
\subsection{Model Architecture}
The Virchow2 model \cite{zimmermann2024virchow2scalingselfsupervisedmixed}, based on the Vision Transformer (ViT) architecture, serves as a powerful feature extractor owing to its multi-head self-attention mechanisms and capacity for learning deep, hierarchical representations. For binary classification, we modified its final classification head to a single output neuron. To enable efficient fine-tuning while reducing computational overhead, we employed Low-Rank Adaptation (LoRA). Specifically, we set the rank to 
r=8 and the scaling factor to  $\alpha=16$, striking a balance between model adaptability and parameter efficiency. A dropout rate of 0.3 was applied to the LoRA modules to enhance regularization. LoRA was integrated into key components of the transformer, including the \texttt{qkv} projections, the output projection layer, and both fully connected layers (\texttt{fc1} and \texttt{fc2}), ensuring minimal modification to the backbone architecture while significantly reducing the number of trainable parameters.

\subsection{Dataset and Split Strategy}
Training leverages annotated mitotic patches from MIDOG 2025 Atypical Set \cite{weiss_2025_15188326},AMi‑Br Breast Cancer dataset \cite{palm_histologic_2025}, Zhuoyan Shen et al. dataset \cite{shen2025omg} and Mitosis subtyping dataset \cite{jahanifar2025mitosis}. All images are resized to 224×224 pixels. To avoid slide-level data leakage, we apply a StratifiedGroupKFold approach (5 folds), dividing data by slide identifiers while balancing atypical vs normal labels across folds.

\subsection{Preprocessing \& Augmentation}
We employ rich augmentations to enhance model generalization:
\begin{itemize}
    \item Random Resized Crop (0.8–1.0 scale), random flips, rotations up to 30°, color jitter (brightness, contrast, saturation, hue), grayscale conversion, random erasing.
    \item  StrongAugment selectively applies transformation sequences to diversify training examples.
    \item Validation transforms are limited to resizing, center cropping, and normalization to simulate inference conditions.

\end{itemize}

\subsection{Training Configuration}
Key hyperparameters include:
\begin{itemize}
    \item  Adam optimizer with learning rate \(5\times10^{-5}\), weight decay \(10^{-5}\).
    \item  Focal Loss with \(\alpha=0.25\) and \(\gamma=2.0\), emphasizing tougher minority class examples.
    \item   WeightedRandomSampler inversely weights class frequencies to counter label imbalance.
    \item Early stopping invoked after 10 consecutive epochs of no validation improvement.
    
\end{itemize}


\subsection{ TTA Inference}
After training each fold, we merge LoRA weights into the base network to produce a standard model. Inference comprises:

1. Test‑Time Augmentation (TTA): apply scaling (0.9, 1.0, 1.1), flips, 90° rotations, brightness adjustments.
2. Predictions averaged across TTA variants and models for robust final classification.

\section{Results}
Per-domain performance on validation folds is reported in Table 2.

\begin{table}[ht]
\centering
\scriptsize
\caption{Domain-wise validation performance}
\begin{tabular}{lccccc}
\toprule
Domain & AUC & Acc & Sens & Spec & Bal. Acc \\
\midrule
0 & 0.820 & 0.750 & 0.500 & 0.781 & 0.641 \\
1 & 0.954 & 0.851 & 0.966 & 0.826 & 0.896 \\
2 & 0.985 & 0.864 & 1.000 & 0.809 & 0.905 \\
3 & 1.000 & 0.921 & 1.000 & 0.917 & 0.958 \\
\midrule
\textbf{Overall} & \textbf{0.964} & \textbf{0.853} & \textbf{0.958} & \textbf{0.827} & \textbf{0.892} \\
\bottomrule
\end{tabular}
\end{table}

Our approach yields highly competitive overall performance. Domains with abundant atypical examples (e.g., Domain 3) reach near-perfect balanced accuracy, whereas more challenging domains (e.g., Domain 0) yield moderate performance, highlighting opportunities for domain adaptation or additional data.

\section{Discussion}
MitoDetect++ demonstrates that LoRA effectively reduces training overhead while preserving performance. Group-aware splits ensure that evaluations reflect true generalization, not memorization of slide-specific patterns. TTA and ensembling substantially improve robustness against domain shifts and input variability.

Domain 0’s relatively lower sensitivity reflects the need for targeted strategies e.g. domain-specific augmentation or focal loss tuning—to handle sparse atypical instances more effectively. Future work could explore domain adversarial training or few-shot learning to address these shortcomings.

\section{Conclusion}
We present MitoDetect++, a resource-efficient, robust pipeline for detecting atypical mitoses in histopathological images. By integrating LoRA-based fine-tuning, strong augmentation, and TTA-ensemble inference within a rigorously validated framework, our method achieves a balanced accuracy of 0.892. Its efficiency and performance make it a strong candidate for clinical and research deployment.

\section*{Acknowledgements}
We thank the MIDOG 2025 organizers and dataset contributors. This work was supported by the Tissue Image Analytics Centre at the University of Warwick.

\bibliographystyle{plain}
\bibliography{main}
\end{document}